\documentclass[twocolumn,preprintnumbers,secnumarabic,amsmath,amssymb,nofootinbib,floatfix]{revtex4}
\usepackage{varioref,exscale,latexsym,amsmath,amssymb}
\usepackage{graphicx}
\usepackage{graphicx}% Include figure files
\usepackage{dcolumn}% Align table columns on decimal point
\usepackage{bm}% bold math

\def\beq{\begin{equation}}

\def\eeq{\end{equation}}

\def\beqa{\begin{eqnarray}}

\def\eeqa{\end{eqnarray}}

\begin{document}

\title{{\bf On Regge kinematics in SCET \\}}

\medskip\
\author{John F. Donoghue}%
\email[Email: ]{donoghue@physics.umass.edu}
\affiliation{Department of Physics,
University of Massachusetts\\
Amherst, MA  01003, USA\\}
\affiliation{Institut f\"{u}r Theoretische Physik\\
Universit\"{a}t Z\"{u}rich\\
8057 Z\"{u}rich, Switzerland}
\author{Daniel Wyler}
\email[Email: ]{wyler@physik.unizh.ch}
\affiliation{Institut f\"{u}r Theoretische Physik\\
Universit\"{a}t Z\"{u}rich\\
8057 Z\"{u}rich, Switzerland}

\begin{abstract}
We discuss the kinematics of the particles that make up a Reggeon in field theory, using the terminology of the Soft Collinear Effective Theory (SCET).
Reggeization sums a series of strongly-ordered collinear emissions resulting in an overall Reggeon exchange that falls in the Glauber or Coulomb kinematic region. This is an extremely multi-scale problem and appears to fall outside of the usual organizing scheme
of SCET.
\end{abstract}
\maketitle

%%%%%%%%%%%%%%%%%%%%%%%%%%%%%%%%%

\section{Introduction}

In the 1960's it was discovered, through a combination of general principles and
phenomenology, that the high energy behavior of hadron scattering amplitudes is
governed by Regge exchanges, in particular by Pomeron exchange\cite{Eden, Eden:1971jm, Collins:1971ff, Collins:1977, Irving:1977ea}. The region of applicability for these techniques is
\begin{equation}
s \to \infty~~~~~~~t ~{\rm fixed}
\end{equation}
where $s,t$ are the usual Mandelstam variables.
While it is known how Regge behavior emerges in a
field theory, the original applications were not derived from any fundamental theory of the
strong interactions. With the advent of QCD, Regge behavior has also been found for
quarks and gluons\cite{Lipatov:1976zz, Frankfurt:1976ds, Tyburski, Kuraev:1976ge, BFKL, Lipatov:1985uk, Fadin:1998py, Forshaw, Donnachie:2002en}. These are applicable in the Regge region as long as $t$
is large enough that a perturbative treatment is possible. This has led to concepts such
as the ``Reggeized gluon'' and the ``hard Pomeron''.
Most recently, these ideas have been revived in the experiments
at HERA\cite{Hera}. While the ``hard'' and ``soft'' Reggeon regions  are
phenomenologically distinct, both types of behavior exist\footnote{Note that the word "soft" in traditional hadronic usage differs somewhat from the same word in the Soft Collinear theory, which has a more technical definition described below.}.

In a more recent  development, an effective field theory for the QCD interactions
of very high energy quarks and gluons has been formulated. This separates the
high energy degrees of freedom interacting with a high energy particle into collinear
modes and soft modes, hence the name Soft Collinear Effective Theory (SCET)\cite{Bauer:2000yr, Beneke:2002ph, Bauer:2001yt, Stewart:2003gt, Neubert:2005mu, Fleming:2009fe}. This theory has been very
useful in organizing theoretical calculations, especially in the decays of heavy hadrons and high
energy phenomena.

If the high energy behavior of
scattering amplitudes is dominated by Regge exchange, one should expect that these ideas must
also find a description within a theory such as SCET that describes the high energy degrees of freedom.
To this end we here discuss
the ideas of Regge theory in the language of SCET. There must be a region of compatibility of these
two approaches. It turns out that a Reggeon exchange is an unusual object in
SCET, one which emerges from a ordered series of collinear exchanges but which produces an object of a different character.

The relevance of Regge exchange for phenomenology follows from the emergence of {\it power-law} behavior
for scattering amplitudes. An amplitude that behaves like $(\alpha_s \ln s)^n$ in some order of perturbation theory
sums to $s^{\alpha (t)}$ in the Regge region. Note that the traditional notation for a Regge exponent $\alpha(t)$
should not be confused with the QCD coupling $\alpha_s$. The latter will always carry the subscript $s$ in this paper.
Since Reggeized gluons and the hard Pomeron (and also the soft Pomeron)
 carry $\alpha(t)>0$, this can lead to an enhanced power behavior of interactions
in the Regge region, with an effect larger than estimated in naive perturbative power counting.

\section{Kinematic regions}

\begin{figure}[ht]
 \begin{center}
  \includegraphics[scale=0.8]{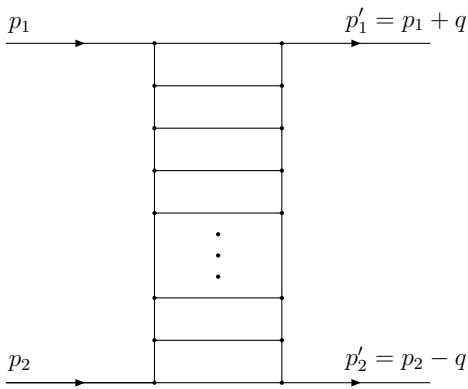}
 \end{center}
 \caption{\small{The ladder graphs }}
 \label{ladder}
\end{figure}

\begin{figure}[ht]
 \begin{center}
  \includegraphics[scale=0.8]{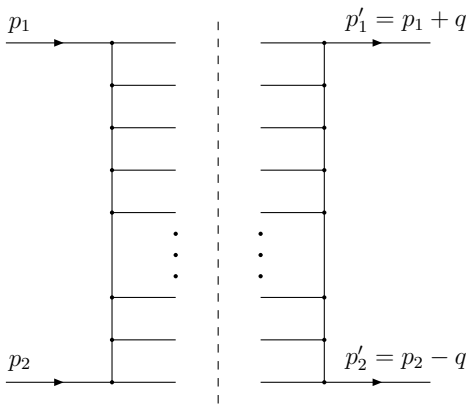}
 \end{center}
 \caption{\small{The cut ladder graphs }}
 \label{cut}
\end{figure}

The asymptotic Regge behavior comes from the summation of the ladder graphs shown in Figure \ref{ladder}. In this figure we treat all particles as scalars because we are primarily interested in kinematics, although we will refer to the particles as gluons.

These ladder diagrams can be interpreted in
a couple of ways. If one is concerned with the net effect of the ladder sum, one sees it as the t channel exchange of
a single object - a Reggeon. Alternatively we can reconstruct the kinematic behavior of the ladder sum from consideration of the
discontinuities in the diagrams, and the relevant discontinuities turn out to be those where the cut lines are the rungs of
the ladder, as in Fig \ref{cut}. In this picture the Reggeon corresponds to the emission of on-shell particles in the s-channel.

We describe the scattering process $p_1+p_2\to p_1' + p_2'$, where as usual $s=(p_1+p_2)^2$, $t=(p_1-p_1')^2$ and $u=(p_1-p_2')^2$.
Defining a right-moving light-like four-vector $n^\mu=(1,0,0,1)$ and a left-moving one $\bar{n}^\mu=(1,0,0,-1)$, we display momenta in light-cone coordinates $p=(p^+,p^-,{\mathbf p_\bot})$,
with $p^+ =n_\mu p^\mu, ~p^- = \bar{n}_\mu p^\mu$. In these coordinates an invariant mass is $p^2 =p_+p_- - {\mathbf p_\bot}^2$. We treat particle 1
as right-moving in the center of mass, and particle 2 as left-moving,
\begin{equation}
p_1 = (p_1^+, 0,0)~~~~~~~p_2=(0,p_2^-,0)
\end{equation}
with $p_1^+=p_2^-=\sqrt{s}$.

As mentioned, the kinematic region of Regge theory is $s \to \infty,~t ~{\rm fixed}$. Let us define a small parameter
\begin{equation}
\lambda = \sqrt{-t/s}~~.
\end{equation}
By definition, the Reggeon then has momentum
\begin{equation}
q^2 = -s \lambda^2
\end{equation}
The individual components can be found from the on-shell condition for the final particles
\begin{eqnarray}
(p_1+q)^2 &=&0  ~~\Rightarrow ~~ (p_1+q)^+q^- -\mathbf{q}_\bot^2 =0 \nonumber \\
(p_2-q)^2&=&0  ~~\Rightarrow  -(p_2-q)^-q^+ -\mathbf{q}_\bot^2 =0
\end{eqnarray}
This implies that the components of the Reggeon are of order
\begin{equation}
q\sim \sqrt{s}(\lambda^2,\lambda^2,\lambda)
\label{glauber}
\end{equation}
Since the only principle used in this relation was the kinematics of the external particles, we can see an important and general feature. To connect an energetic right-moving particle with an energetic left-mover, with limited momentum exchange, one requires an exchanged particle which is dominated by the transverse momentum. This kinematic configuration has been called either Coulombic or Glauber\cite{glauber}, as it is typical of the region of Coulomb exchange at very high energies.

In SCET one also classifies the momenta by powers of a small scale. The standard names and scaling properties for these are
\begin{eqnarray}
{\rm Collinear}\Rightarrow ~~~p &\sim & \sqrt{s}(1, \lambda^2,\lambda),~~~~~~~p^2\sim s \lambda^2 \nonumber \\
{\rm Soft}\Rightarrow~~~~ ~~p &\sim & \sqrt{s}(\lambda, \lambda,\lambda),~~~~~~~~ p^2\sim s\lambda^2 \nonumber \\
{\rm Ultra-soft}\Rightarrow~~~p&\sim & \sqrt{s}(\lambda^2, \lambda^2,\lambda^2), ~~~~p^2\sim s \lambda^4
\end{eqnarray}
Hardly studied in SCET is the Glauber region\cite{glauber2}
\begin{equation}
{\rm Glauber}\Rightarrow~~~~~p\sim \sqrt{s}(\lambda^2, \lambda^2,\lambda), ~~~~~~~p^2\sim s\lambda^2
\end{equation}
The Glauber region can be considered as an endpoint of the soft region in which the longitudinal momenta are particulary small, but the key feature describing the region is that the dominant component is the transverse momentum, which itself is of order $\lambda$. There does not exist a consistent treatment of this region yet.

In ${\rm SCET}_I$, the small parameter is taken as
\begin{equation}
\lambda_I = \sqrt{\Lambda/Q}
\end{equation}
with $Q$ being a large scale, and in ${\rm SCET}_{II}$, the parameter is
\begin{equation}
\lambda_{II} = \Lambda/Q
\end{equation}
Since $Q\sim \sqrt{s}$, our parameter is close to $\lambda_{II}$, if $\Lambda \sim \sqrt{-t}$. (Note however that $\sqrt{-t}$ can still be much larger than the scale of QCD, $\Lambda_{QCD}$.)

While the overall Reggeon exchange corresponds to an object which is in the Glauber kinematic regime, we will see that the components that make up the legs and rungs of the ladder diagrams in general do not fit this classification, but are rather an ordered set of mostly collinear particles. Within this ordering scheme there is another small parameter which differentiates the ladder rungs from ordinary collinear particles. This extra ordering parameter is at present not included in the SCET classification.

\section{Kinematics of internal lines in the ladder sum}

In this section we review the field theory treatment of Regge exchange in order to make the connection to SCET. Our interest is in understanding the nature of the gluon kinematics for the internal lines within the ladder sum.  The derivations of this section are well known in the Regge community and can be found in standard texts. The novel aspect is the rephrasing of the results within the context of SCET.

Regge behavior is readily found in field theory. The ladder diagrams in general will depend on both $s$ and $t$. As $s \to \infty$, only a tiny corner of the loop integration region will give the leading contribution. In this region the $n$-loop diagram yields
\begin{equation}
\frac{g^2}{s~n!} [\beta(t)\ln (-s)]^{n}
\end{equation}
\begin{equation}
\beta (t) = -\frac{g^2}{16\pi^2}\int dy_1 dy_2 \frac{\delta (1-y_1-y_2)}{[y_1y_2 t -m^2(y_1+y_2)]}
\end{equation}
Here $g$ is the coupling constant at each vertex in the ladder diagram.
Summing over all possible ladder diagrams leads to the result
\begin{equation}
\frac{g^2}{s} \sum_{n=0}^\infty \frac{1}{n!}[\beta(t)\ln (-s)]^{n}  = g^2 s^{\alpha(t)}
\end{equation}
with the Regge exponent
\begin{equation}
\alpha (t) = -1 + \beta(t)
\end{equation}
The power dependence of the sum is significantly different from that of any one of
the individual terms

The Regge exponent can also be written as a two dimensional integral
\begin{equation}
\beta (t) = -\frac{g^2}{16\pi^2}\int d^2k \frac{1}{[\mathbf{k}^2+m^2][(\mathbf{k-q})^2 +m^2]}
\label{transverse}
\end{equation}
The transverse momentum in this integral is of order $\sqrt{|t|}$, for $|t|>m^2$.
The propagators for the horizontal rungs do not appear in this result and the asymptotic behavior can be
found from a ``pinched'' graph where horzontal lines are ommitted. The factor $\beta(t)$ corresponds
to the {\em two-dimensional} integral over the transverse momenta of the vertical legs.

\subsection{Discontinuities of the ladder diagram}

It is easier to understand the kinematics internal to the ladder sum if we use the fact that the
leading behavior can be reconstructed from the discontinuity in the ladder diagram\cite{Chang:1971je, Forshaw}. As usual the relationship is
\begin{equation}
Amp \sim \frac{1}{n!} [\ln (-s)]^n   ~~~\Leftrightarrow~~~Disc \sim \frac{\pi}{(n-1)!}[\ln(s)]^{n-1}
\end{equation}
so that the discontinuity will contain one less power of $\ln( s)$.
There is a thorough treatment of this approach in the fine lecture notes of Forshaw and Ross\cite{Forshaw}.

\begin{figure}[ht]
 \begin{center}
 \includegraphics[scale=0.7]{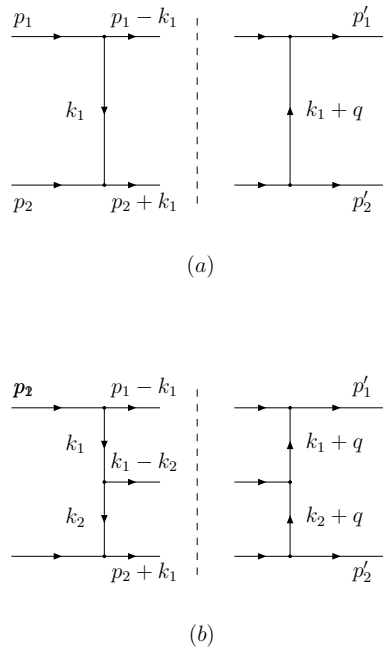}
 \end{center}
 \caption{\small{The discontinuity of the box diagram and of the three-rung diagram.}}
 \label{cutbox}
\end{figure}

The discontinuity puts all the particles across the cut on their mass shell. The
imaginary part of the diagram comes then directly from this on-shell configuration and the real part comes from
particles in the neighborhood of the on-shell point. The relevant discontinuities of the box diagram and the three-rung ladder are shown in Fig. \ref{cutbox},
while the general diagram was previously shown in Fig \ref{cut}.

For the box diagram of Figure \ref{cutbox}a, it is immediately clear from the argument given in Sec 2 that the exchanged gluons fall in the Glauber or Coulombic regime. In each half of the diagram, the exchanged particle connects an on-shell right moving particle to an on-shell left mover. By the argument of Sec. 2 the exchanged particle then must satisfy Eq. \ref{glauber}.
However, this kinematic condition is not common to all the ladder diagrams, as we will see next.

The first non-trivial example - the three rung
ladder of Fig. \ref{cutbox}b displays what turns out to be the typical pattern in the ladder sum.

 Let us label the exchanged gluon momenta - the vertical leg - with an extra minus sign
convention
\begin{equation}
 k_i = (k_i^+, -k_i^-,\mathbf{k_i}_\bot)
\end{equation}
This convention is connected to the choice to draw the momentum flowing down the ladder rather than up - it will save many minus
signs and reversed inequalities in what follows.
The discontinuity involves the integrations over the available phase
space plus the on-shell constraints, specifically\cite{Forshaw}
\begin{eqnarray}
Disc &\sim& \int d^4k_1d^4k_2 \nonumber \\
&\times&\delta[(p_1-k_1)^2] \delta[(p_2+k_2)^2]\delta[(k_1-k_2)^2]  ~P(k,q)  \nonumber \\ \\
&\sim&\int d^2k_{1\bot}d^2k_{2\bot}\int dk_1^+dk_1^-dk_2^+dk_2^- \nonumber \\ &\times&\delta[(p_1-k_1)^+k_1^--\mathbf{k}_{1\bot}^2] \delta[(p_2-k_2)^-k_2^+-\mathbf{k}_{2\bot}^2] \nonumber \\
&\times& \delta[ (k_1-k_2)^+ (k_2-k_1)^- -\Delta \mathbf{k}_\bot^2]~P(k,q) \nonumber
\label{disc}
\end{eqnarray}
where we have dropped the
masses in the delta function constraints, and where the propagators are
\begin{equation}
P(k,q) = \frac{1}{k_1^2-m^2}\frac{1}{k_2^2-m^2}\frac{1}{(k_1-q)^2-m^2}\frac{1}{(k_2-q)^2-m^2}
\end{equation}
We can see from the delta-functions that the on-shell constraints require an ordering of the momenta
\begin{eqnarray}
p_1^+>k_1^+>k_2^+>p_2^+&=&0  \nonumber \\
p_2^->k_2^->k_1^->p_1^-&=&0
\end{eqnarray}
At this stage, the delta-functions enforcing the on-shell constraint for the upper and
lower legs can be performed
using the $k_1^- ,k_2^+ $ variables in order to obtain
\begin{eqnarray}
Disc &\sim& \int d^2k_{1\bot}d^2k_{2\bot}\int^{\sqrt{s}} dk_1^+dk_2^- \nonumber \\
&\times&
 \delta[ (k_1-k_2)^+ (k_2-k_1)^- -\Delta \mathbf{k}_\bot^2]~P(k,q)
\end{eqnarray}
with
\begin{equation}
k_1^- = \frac{\mathbf{k_1}_\bot^2 }{\sqrt{s}-k_1^+}  ~~;~~k_2^+= \frac{\mathbf{k_2}_\bot^2 }{\sqrt{s}-k_2^-}
\end{equation}

After these simple constraints, the interesting remaining constraint follows from the remaining integration of the middle rung of the diagram
\begin{equation}
\int^{\sqrt{s}} dk_1^+~dk_2^-~\delta[(k_1-k_2)^+ (k_2-k_1)^--\Delta\mathbf{k}_\bot^2] P(k_1,k_2)
\end{equation}
We would like to demonstrate that the region of Regge exponentiation comes from the region of strong ordering
\begin{eqnarray}
p_1^+&>>&k_1^+  \nonumber \\
p_2^-&>>&k_2^- \label{strongorder}
\end{eqnarray}
which then forces the other components to be highly suppressed
\begin{equation}
k_1^- = \frac{\mathbf{k_1}_\bot^2}{\sqrt{s}} \sim \lambda^2 \sqrt{s} ~~;~~k_2^+= \frac{\mathbf{k_2}_\bot^2 }{\sqrt{s}}\sim \lambda^2\sqrt{s} \label{small}
\end{equation}
such that strong ordering holds throughout with
\begin{eqnarray}
p_1^+>>k_1^+>>k_2^+>>p_2^+&=&0  \nonumber \\
p_2^->>k_2^->>k_1^->>p_1^-&=&0
\end{eqnarray}
We will give a physical interpretation of this strong ordering later.

\subsection{The kinematics of strong-ordering}

In order to demonstrate the origin and effect of strong-ordering of $k_1$ and $k_2$,
let us break the integration region up into four parts:\\
\noindent 1) {\bf Both strongly ordered}
\begin{equation}
k_1^+<\eta \sqrt{s},~~~~~~~k_2^-<\eta \sqrt{s}
\end{equation}
\noindent 2) {\bf One strongly ordered}
\begin{equation}
 \eta \sqrt{s}< k_1^+ <\sqrt{s},~~~~~~~~~~k_2^-<\eta \sqrt{s}
\end{equation}
\noindent 3) {\bf One strongly ordered}
\begin{equation}
k_1^+<\eta \sqrt{s},~~~~~~~~~~  \eta \sqrt{s}<k_2^-<\sqrt{s}
\end{equation}
\noindent 4) {\bf Neither strongly ordered}
\begin{equation}
\eta \sqrt{s}<k_1^+< \sqrt{s},~~~~~~~\eta \sqrt{s}<k_2^-< \sqrt{s}
\end{equation}
where $\eta$ is a small number independent of $s$.

Region 1 is the one with strong
ordering as in Eq. \ref{strongorder}. In it, Eq. \ref{small} holds for the small momenta. This
implies that all of the remaining propagators in $P(k_1,k_2)$ are largely transverse. For example,
\begin{eqnarray}
k_1^2-m^2 &=& -k_1^+k_1^- -\mathbf{k}_{1\bot}^2-m^2 \\
&\approx& -\eta\mathbf{k}_{1\bot}^2- \mathbf{k}_{1\bot}^2-m^2 \approx -(\mathbf{k}_{1\bot}^2+m^2) \nonumber
\label{transprop}
\end{eqnarray}
i.e. the longitudinal components can be neglected and the propagator is transverse. In this case the
constraint integral can be easily done
\begin{eqnarray}
I_1 &=& \int^{\eta\sqrt{s} } dk_1^+~dk_2^-~\delta(k_1^+ k_2^--(\mathbf{k}_{1\bot}^2- \mathbf{k}_{2\bot}^2)) \nonumber \\
&=&\int_{\Delta\mathbf{k}_\bot^2/\sqrt{s}}^{\eta\sqrt{s} } \frac{dk_1^+}{k_1^+} \nonumber \\
&=& \ln (\frac{s}{\eta^2(\mathbf{k}_{1\bot}^2- \mathbf{k}_{2\bot}^2)})\to  \ln(s)
\end{eqnarray}
where the final limit is valid as $s\to \infty$.
Completing the calculation of the discontinuity in this region then leads to
\begin{eqnarray}
&~&Disc \sim \ln(s) \int d^2k_{1\bot}d^2k_{2\bot} \\
&\times& \frac{1}{\mathbf{k}_{1\bot}^2-m^2}\frac{1}{\mathbf{k}_{2\bot}^2-m^2}\frac{1}{(\mathbf{k}_1-\mathbf{q})_\bot^2-m^2}
\frac{1}{(\mathbf{k}_2-\mathbf{q})^2-m^2}\nonumber
\label{factored}
\end{eqnarray}
which factors into the product of two transverse integrations of the form seen previously in the
field theory
calculation, i.e. Eq. \ref{transverse}. Adding in coupling constants and the appropriate numerical factors
leads to agreement with the field theory result for the leading structure of this loop amplitude.

Region 2 (and symmetrically in Region 3) does not contribute to the leading approximation.
The longitudinal momentum does not decouple from the propagator and there is no factor of $\ln s$ which
develops. Even ignoring the propagator factors we would have
\begin{eqnarray}
I_1 &=&\int_{\eta\sqrt{s}}^{\sqrt{s} } \frac{dk_1^+}{k_1^+} \nonumber \\
&=& -\ln \eta
\end{eqnarray}
This contains no factor of $\ln s$.

In Region 4, the delta function constraint cannot be satisfied since $\eta^2 s>>(\mathbf{k}_{1\bot}^2- \mathbf{k}_{2\bot}^2)$.

The strong ordering generalizes to the general ladder diagram, Fig. 2. For all of the cut lines to be on-shell requires
\begin{eqnarray}
p_1^+>k_1^+>k_2^+>.....>k_n^+>p_2^+&=&0  \nonumber \\
p_2^->k_n^->.....>k_2^->k_1^->p_1^-&=&0
\end{eqnarray}
However again the main contribution comes from the strong ordered regions
\begin{eqnarray}
p_1^+>> k_1^+>>k_2^+>.....>>k_n^+>>p_2^+&=&0  \nonumber \\
p_2^->> k_n^->>.....>>k_2^->>k_1^->>p_1^-&=&0
\label{strongorder2}
\end{eqnarray}

This notion of strong ordering is perhaps the key feature of the Regge kinematic regime. The main effect is to allow an approximation to the scattering amplitude. If one considers the general denominator of the propagator which is displayed in the first line of Eq. 31, we see that the form that gets exponentiated in the Regge sum arises only when the longitudinal product $k^+k^-$ can be neglected in comparison to the transverse momentum. This is only in the strongly ordered region. This yields further simplifications in the overall momentum integration.
Because regions 2 and 3 do not contribute to the leading $\ln (s)$ enhancement, one is free to
approximate the propagators by dropping the longitudinal momenta as in the second line of Eq. 31 in these regions. The propagators than have the factored form of
Eq. \ref{transverse}, and the full integration then can be done
simply. In the example above, the key result of the strong ordering is to allow the propagators to be
approximated by their transverse components only, as in Eq. 15, 33.

\section{The hierarchy of ordered collinear gluons}

Within SCET, the ladder diagrams become a true multi-scale problem. The strong ordering condition means that there is a continuum of
scales from the largest scale $s$ down to the smallest scale $t$. Let us label the magnitude of the strong
ordering by the factor $\eta$, i.e. $k_{i+1}^+\sim \eta k_{i}^+$ and $k_{i-1}^-\sim \eta k_i^-$. There is no unique
value of $\eta$ that one must use. Since, as explained above, the primary use of strong ordering is to allow an approximation
to the amplitude, for example to neglect $ k_{i+1}^+$ compared to $k_i^+$, the value of $\eta$ is related to the accuracy of the
Regge result. Certainly $\eta=1/10$ allows an accurate approximation, while $\eta=1/3$ is not as good. Additionally, since the
momentum in the legs is integrated over, even within one diagram there is not a fixed value of $\eta$. However, we will use this
parameter as an indicator that the dominant contributions come from regions which obey the strong ordering condition.
Presumably the fact that
the Regge region of high energy scattering begins at c.m. energies somewhat above $1$~GeV is related to the condition that $s$ has to be
far enough above the average $t$ so that the ladder approximation - with many rungs due to the strong ordering - is valid.

For a concrete realization of what is implied by the strong ordering condition, let us look at an extreme example. Let us take
$\sqrt{s} =10^5$ in GeV units and consider a 11 rung diagram. If we have momentum ordered by a fixed factor of $\eta = 1/10$, then the
momenta of the legs of the ladder (labeled by $k_i$) have the following ordering (still in GeV units):
\begin{eqnarray}
p_1 &=& 10^5 n \nonumber \\
k_1 &=& 10^4 n + 10^{-5} \bar{n} + k_\bot  \nonumber \\
k_2 &=& 10^3 n + 10^{-4} \bar{n} + k_\bot  \nonumber \\
k_3 &=& 10^2 n + 10^{-3} \bar{n} + k_\bot  \nonumber \\
k_4 &=& 10^1 n + 10^{-2} \bar{n} + k_\bot  \nonumber \\
k_5 &=& 10^0 n + 10^{-1} \bar{n} + k_\bot  \nonumber \\
k_6 &=& 10^{-1} n + 10^{0} \bar{n} + k_\bot  \nonumber \\
k_7 &=& 10^{-2} n + 10^{1} \bar{n} + k_\bot  \nonumber \\
k_8 &=& 10^{-3} n + 10^{2} \bar{n} + k_\bot  \nonumber \\
k_9 &=& 10^{-4} n + 10^{3} \bar{n} + k_\bot  \nonumber \\
k_{10} &=& 10^{-5} n + 10^{4} \bar{n} + k_\bot  \nonumber \\
p_2 &=& 10^5 \bar{n}
\end{eqnarray}
The rungs of the ladder satisfy $r_i = k_i-k_{i+1}$. In this example, we see a transition from right-moving collinear gluons near the top rung to left-moving particles near the bottom.
Each is dominatingly transverse since $k_i^2 =-\mathbf{k}_\bot^2 + 10^{-1}~{\rm GeV}$.
There are two legs ($k_{5,6}$) that are considered in the Glauber regime, in which the transverse momentum is the largest component and both longitudinal momenta are sub-dominant. It is such a leg that allows the connection of the right-moving particles in the upper portion of the diagram with the left-moving branch in the lower portion. Of course, this example is incomplete
since the $k_i^\pm$ momenta are really integrated over rather than being in a fixed ratio.
However it does illustrate an essential feature of strong
ordering - there is a transition of right-collinear through at least one Glauber leg to left-collinear within any given ladder diagram.

The physical interpretation of this is that we have a series of ordered and nearly collinear emissions. The rungs of the
latter can be taken to be on-shell, but the general linear momenta is still high such that the collinear pinch is applicable.
The vertical legs of the ladder are off-shell by order $t/s$, but also carry a high linear momentum along the $n$ or $\bar{n}$
direction. The legs near the top of the ladder are dominantly right-moving, those near the bottom are dominantly left-moving.
Somewhere in the chain there is a Glauber leg connecting these.  The whole ordered set of diagrams sums to produce an net effect that is equivalent to a
Glauber or Coulomb exchange.

When we reformulate this process in the language of SCET we encounter a difficulty with the
multiple scales. In standard SCET ordering one considers a small parameter $\lambda$ and orders
the effective field theory in powers of $\lambda$. Here, however, we have at least two small
parameters. We can take one of these to be $\lambda\sim \sqrt{-t/s}$ which is treated as an extremely small
parameter in the Regge limit. The other is what we have called $\eta$ above, i.e. the strong ordering parameter.
This parameter is independent of $s,t$ and clearly $\eta>>\lambda\approx 0$ in the Regge region.
If we consider a vertex along one of the legs of a right-moving legs neat the top of a ladder diagram, the relevant ordering of the momenta is
\begin{eqnarray}
k_i &=& \sqrt{s_i}(1,0,0) \nonumber \\
k_{i+1} &=&  \sqrt{s_i}(\eta,\sim 0,\lambda) \nonumber \\
r_i &=& k_i-k_{i+1} \sim k_i + O(\lambda)
\end{eqnarray}
with $\lambda\sim \sqrt{-t/s_i}$, $\lambda<<\eta<<1$ and $s_i \sim O(\eta^i s)>> \Lambda_{QCD}^2$. In words, we would label particle $r_i$ as being collinear to $k_i$. Because of the two small parameters, particle $k_{i+1}$
does not have a standard name in SCET. We can give it a name of ``ordered collinear'' to imply that it shares the directions of the original gluon, yet is at a parameterically smaller momentum. That is:
\begin{equation}
{\rm ordered ~collinear}\Rightarrow~~~~~p\sim \sqrt{s}(\eta, \lambda^2,\lambda), ~~~~~~~p^2\sim s\lambda^2
\end{equation}
with $\lambda<<\eta<<1$.
When one has a whole ladder of such vertices, one has a highly multiscale problem. If the exchange particles are treated as regular collinear particles, the ordered region is at an endpoint of the integration over the longitudinal momentum, but properties derived for collinear particles in general will not be appropriate for this endpoint. In particular, a treatment of general collinear particles will miss the Regge kinematic region unless care is taken to properly treat the strongly ordered region and also to include the potential for a Glauber exchange connecting left-movers and right-movers.

Even though most of the particles in the ladder diagram are of the ordered collinear variety, the net effect of a Reggeon exchange falls in the Glauber or Coulombic kinematic regime.

\section{Summary }

The application of Regge ideas to QCD has become an extensive subfield. The
main results are that there appears a Reggeon with the quantum numbers of the
gluon - the Reggeized gluon\cite{Lipatov:1976zz, Frankfurt:1976ds, Tyburski} - as well as a hard Pomeron\cite{BFKL} with vacuum quantum numbers. The Reggized gluon is not
just the ``dressing'' of a gluon, but is
actually built out of {\it two} gluon exchange in a spin-one color octet channel.
The hard Pomeron is then built out of
the exchange of two Reggeized gluons, and is a color singlet object.

The power-law behavior of Regge scattering has the possibility to influence the phenomenology of
high energy processes.
An example that already exists in the literature is the work by DGPS\cite{DGPS} on
final state interactions in B meson decay. In this work, the region of soft final
state interaction was shown to be of order $1/m_B^2$; however, the exchange of a soft Pomeron
provides a scattering amplitude which grows as $s^{1.04} \sim (m_B^2)^{1.04}$ which removes the
nominal suppression. While this study was phenomenological in nature and invoked the soft Pomeron,
it provides the archetype for how Regge phenomena can modify the power counting of perturbative
calculations\cite{BBNS}.

To proceed further, one should calculate directly with the Reggeized gluon and the hard Pomeron. However, the matching of these degrees of freedom with the usual gluons is non-trivial, and we hope to report on this matching in a future publication. This should allow one to place the usual hard Regge phenomenology on a better footing as well as to complete SCET through the inclusion of the Regge region. The matching to soft Regge physics is more difficult and it is not clear that this can be accomplished within SCET.

\section*{Acknowledgments} This work is supported in part by the U.S. National Science Foundation through grant PHY- 055304. John Donoghue thanks the Universit\"{a}t Z\"{u}rich for kind hospitality during the period in which most of this work was accomplished. We thank Martin Beneke and Thomas Gehrmann for valuable discussions during the course of this work.


\begin{thebibliography}{99}

\bibitem{Eden}
R. J. Eden, P. V. Landshoff, D. Olive and J. C. Polkinghorne ``{\it The Analytic S matrix}'' (Cambridge University Press, 1966)

%\cite{Eden:1971jm}
\bibitem{Eden:1971jm}
  R.~J.~Eden,
  ``Regge Poles And Elementary Particles,''
  Rept.\ Prog.\ Phys.\  {\bf 34} (1971) 995.
  %%CITATION = RPPHA,34,995;%%


%\cite{Collins:1971ff}
\bibitem{Collins:1971ff}
  P.~D.~B.~Collins,
  ``Regge theory and particle physics,''
  Phys.\ Rept.\  {\bf 1} (1971) 103.
  %%CITATION = PRPLC,1,103;%%
\bibitem{Collins:1977}
  P.~D.~B.~Collins,
  ``{\it An introduction to Regge theory and high energy physics},''
  (Cambridge University Press,  1977)

%\cite{Irving:1977ea}
\bibitem{Irving:1977ea}
  A.~C.~Irving and R.~P.~Worden,
  ``Regge Phenomenology,''
  Phys.\ Rept.\  {\bf 34} (1977) 117.
  %%CITATION = PRPLC,34,117;%%





%\cite{Lipatov:1976zz}
\bibitem{Lipatov:1976zz}
  L.~N.~Lipatov,
  ``Reggeization Of The Vector Meson And The Vacuum Singularity In Nonabelian
  Gauge Theories,''
  Sov.\ J.\ Nucl.\ Phys.\  {\bf 23} (1976) 338
  [Yad.\ Fiz.\  {\bf 23} (1976) 642].
  %%CITATION = YAFIA,23,642;%%

\bibitem{Frankfurt:1976ds}
  L.~L.~Frankfurt and V.~E.~Sherman,
  ``Reggeization Of Vector Meson And Vacuum Singularity In Renormalizable
  Yang-Mills Models,''
  Sov.\ J.\ Nucl.\ Phys.\  {\bf 23} (1976) 581
  %%CITATION = SJNCA,23,581;%%


\bibitem{Tyburski}
L. Tyburski, ``Reggeization of the fermion-fermion scattering amplitude in non-Abelian gauge theories'', Phys. Rev. {\bf D13} (1976) 1107.



\bibitem{Kuraev:1976ge}
  E.~A.~Kuraev, L.~N.~Lipatov and V.~S.~Fadin,
  ``Multi - Reggeon Processes In The Yang-Mills Theory,''
  Sov.\ Phys.\ JETP {\bf 44} (1976) 443
  [Zh.\ Eksp.\ Teor.\ Fiz.\  {\bf 71} (1976) 840].
  %%CITATION = ZETFA,71,840;%%




%\cite{Balitsky:1978ic}
\bibitem{BFKL}
  I.~I.~Balitsky and L.~N.~Lipatov,
  ``The Pomeranchuk Singularity In Quantum Chromodynamics,''
  Sov.\ J.\ Nucl.\ Phys.\  {\bf 28} (1978) 822
  [Yad.\ Fiz.\  {\bf 28} (1978) 1597].\\
  %%CITATION = YAFIA,28,1597;%%
E.~A.~Kuraev, L.~N.~Lipatov and V.~S.~Fadin,
  ``The Pomeranchuk Singularity In Nonabelian Gauge Theories,''
  Sov.\ Phys.\ JETP {\bf 45} (1977) 199
  [Zh.\ Eksp.\ Teor.\ Fiz.\  {\bf 72} (1977) 377].
  %%CITATION = ZETFA,72,377;%%



\bibitem{Lipatov:1985uk}
  L.~N.~Lipatov,
  ``The Bare Pomeron In Quantum Chromodynamics,''
  Sov.\ Phys.\ JETP {\bf 63}, 904 (1986)
  [Zh.\ Eksp.\ Teor.\ Fiz.\  {\bf 90}, 1536 (1986)].
  %%CITATION = ZETFA,90,1536;%%


%\cite{Fadin:1998py}
\bibitem{Fadin:1998py}
  V.~S.~Fadin and L.~N.~Lipatov,
  ``BFKL pomeron in the next-to-leading approximation,''
  Phys.\ Lett.\  B {\bf 429}, 127 (1998)
  [arXiv:hep-ph/9802290].
  %%CITATION = PHLTA,B429,127;%%

\bibitem{Forshaw}
  J.~R.~Forshaw and D.~A.~Ross,
  ``{\it Quantum chromodynamics and the pomeron},''
  Cambridge Lect.\ Notes Phys.\  {\bf 9} (1997) 1.
  %%CITATION = 00385,9,1;%%

%\cite{Donnachie:2002en}
\bibitem{Donnachie:2002en}
  S.~Donnachie, H.~G.~Dosch, O.~Nachtmann and P.~Landshoff,
  ``{\it Pomeron physics and QCD},''
  Camb.\ Monogr.\ Part.\ Phys.\ Nucl.\ Phys.\ Cosmol.\  {\bf 19} (2002) 1.
  %%CITATION = CMPCE,19,1;%%



\bibitem{Hera}
  M.~Kapishin  [H1 and ZEUS Collaborations],
  ``Diffraction and vector meson production at HERA,''
  Fizika B {\bf 17} (2008) 131.
  %%CITATION = FZKAA,B17,131;%%
 E.~Iancu, K.~Itakura and S.~Munier,
  ``Saturation and BFKL dynamics in the HERA data at small x,''
  Phys.\ Lett.\  B {\bf 590}, 199 (2004)
  [arXiv:hep-ph/0310338].\\
  %%CITATION = PHLTA,B590,199;%%
J.~Ellis, H.~Kowalski and D.~A.~Ross,
  ``Evidence for the Discrete Asymptotically-Free BFKL Pomeron from HERA
  Data,''
  Phys.\ Lett.\  B {\bf 668}, 51 (2008)
  [arXiv:0803.0258 [hep-ph]].
  %%CITATION = PHLTA,B668,51;%%

\bibitem{Bauer:2000yr}
  C.~W.~Bauer, S.~Fleming, D.~Pirjol and I.~W.~Stewart,
  ``An effective field theory for collinear and soft gluons: Heavy to light
  decays,''
  Phys.\ Rev.\  D {\bf 63}, 114020 (2001)
  [arXiv:hep-ph/0011336].
  %%CITATION = PHRVA,D63,114020;%%

\bibitem{Beneke:2002ph}
  M.~Beneke, A.~P.~Chapovsky, M.~Diehl and T.~Feldmann,
  ``Soft-collinear effective theory and heavy-to-light currents beyond  leading
  power,''
  Nucl.\ Phys.\  B {\bf 643}, 431 (2002)
  [arXiv:hep-ph/0206152].
  %%CITATION = NUPHA,B643,431;%%



\bibitem{Bauer:2001yt}
  C.~W.~Bauer, D.~Pirjol and I.~W.~Stewart,
  ``Soft-Collinear Factorization in Effective Field Theory,''
  Phys.\ Rev.\  D {\bf 65}, 054022 (2002)
  [arXiv:hep-ph/0109045].
  %%CITATION = PHRVA,D65,054022;%%

\bibitem{Stewart:2003gt}
  I.~W.~Stewart,
  ``Theoretical introduction to B decays and the soft-collinear effective
  theory,''
  arXiv:hep-ph/0308185.
  %%CITATION = HEP-PH/0308185;%%


\bibitem{Neubert:2005mu}
  M.~Neubert,
  ``Effective field theory and heavy quark physics,''
  arXiv:hep-ph/0512222.
  %%CITATION = HEP-PH/0512222;%%

%\cite{Fleming:2009fe}
\bibitem{Fleming:2009fe}
  S.~Fleming,
  %``Soft Collinear Effective Theory: An Overview,''
  arXiv:0907.3897 [hep-ph].
  %%CITATION = ARXIV:0907.3897;%%





\bibitem{glauber}
  J.~C.~Collins, D.~E.~Soper and G.~Sterman,
  ``Factorization For Short Distance Hadron - Hadron Scattering,''
  Nucl.\ Phys.\  B {\bf 261}, 104 (1985).\\
  %%CITATION = NUPHA,B261,104;%%
 J.~C.~Collins, D.~E.~Soper and G.~Sterman,
  ``Soft Gluons and Factorization,''
  Nucl.\ Phys.\  B {\bf 308}, 833 (1988).\\
  %%CITATION = NUPHA,B308,833;%%
  S.~M.~Aybat and G.~Sterman,
  ``Soft-Gluon Cancellation, Phases and Factorization with Initial-State
  Partons,''
  Phys.\ Lett.\  B {\bf 671}, 46 (2009)
  [arXiv:0811.0246 [hep-ph]].


\bibitem{glauber2}
  F.~Liu and J.~P.~Ma,
  ``Glauber Gluons in Soft Collinear Effective Theory and Factorization of
  Drell-Yan Processes,''
  arXiv:0802.2973 [hep-ph].\\
 A.~Idilbi and A.~Majumder,
  ``Extending Soft-Collinear-Effective-Theory to describe hard jets in dense
  QCD media,''
  arXiv:0808.1087 [hep-ph].



\bibitem{Polkinghorne}
J. C. Polkinghonre, J. Math. Phys. {\bf 4} (1963) 503, 1393, 1396.


\bibitem{Chang:1971je}
  S.~J.~Chang and T.~M.~Yan,
  ``High-energy elastic and inelastic scattering in phi-to-the-third theory,''
  Phys.\ Rev.\  D {\bf 4}, 537 (1971).
  %%CITATION = PHRVA,D4,537;%%




\bibitem{DGPS}
  J.~F.~Donoghue, E.~Golowich, A.~A.~Petrov and J.~M.~Soares,
  ``Systematics of soft final state interactions in $B$ decay,''
  Phys.\ Rev.\ Lett.\  {\bf 77}, 2178 (1996)
  [arXiv:hep-ph/9604283].
  %%CITATION = PRLTA,77,2178;%%






\bibitem{BBNS}
  M.~Beneke, G.~Buchalla, M.~Neubert and C.~T.~Sachrajda,
  ``{QCD} factorization for $B \to \pi\pi$ decays: Strong phases and CP  violation
  in the heavy quark limit,''
  Phys.\ Rev.\ Lett.\  {\bf 83}, 1914 (1999)
  [arXiv:hep-ph/9905312].
  %%CITATION = PRLTA,83,1914;%%

\bibitem{low}
  F.~E.~Low,
  ``A Model Of The Bare Pomeron,''
  Phys.\ Rev.\  D {\bf 12}, 163 (1975).\\
  %%CITATION = PHRVA,D12,163;%%
  S.~Nussinov,
  ``Colored Quark Version Of Some Hadronic Puzzles,''
  Phys.\ Rev.\ Lett.\  {\bf 34}, 1286 (1975).


\end{thebibliography}
\end{document}